\documentclass[%
 reprint,
]{revtex4-2}

\usepackage{csquotes}
\usepackage[british]{babel}
\usepackage{graphicx}
\usepackage{hyperref}

\usepackage{dcolumn}
\usepackage{bm}
\usepackage[version=4]{mhchem}
\usepackage{enumerate}
\usepackage{enumitem, kantlipsum}
\usepackage{setspace}

\usepackage{enumerate}
\usepackage{xcolor}
\begin{document}

\preprint{APS/123-QED}

\title{Exploring $\beta^+$ decay/EC residues in $^{118}$Sn($^{12}$C,\textit{x})$^{130}$Ba reaction}%

\author{Priyanka$^1$}
\email{priyanka.21phz0015@iitrpr.ac.in}
\author{Amanjot$^1$}
\author{Rupinderjeet Kaur$^1$}
\author{Arshiya Sood$^2$}
\author{Malika Kaushik$^1$}
\thanks{Presently at: National Institute of Technology (NIT) Kurukshetra, Kurukshetra - 136 119, Haryana, India}
\author{Yashraj Jangid$^3$}
\author{Rakesh Kumar$^3$}
\thanks{deceased}
\author{Manoj K. Sharma$^4$}
\author{Pushpendra P. Singh$^1$}
\affiliation{$^1$Department of Physics, Indian Institute of Technology Ropar, Rupnagar - 140 001, Punjab, India}
\affiliation{$^2$DeSiS team - Institut Pluridisciplinaire Hubert Curien- CNRS/Unistra - 67037, Strasbourg, France}
\affiliation{$^3$Inter-University Accelerator Center, New Delhi - 110067, India}
\affiliation{$^4$Physics Department, University of Lucknow, Lucknow - 226 007, Uttar Pradesh, India}

\date{\today}

\begin{abstract}
The fusion cross-sections of $^{126}$Ba, $^{127,126,125}$Cs, $^{125,123,122}$Xe and $^{124,123}$I residues, populated via $x$n, p$x$n, $\alpha$$x$n, and $\alpha$p$x$n channels, have been measured in $^{12}$C+$^{118}$Sn system at E$_{\textrm{lab}}$ $\approx$ 65-85 MeV using offline $\gamma$-spectroscopy. To gain insights into the formation and decay modes of these residues, experimentally measured cross-sections have been analyzed using the statistical model codes PACE4 and EMPIRE. In the analysis, the cross-sections of p$x$n ($^{127,126,125}$Cs), $\alpha$xn ($^{125}$Xe), and $\alpha$p$x$n ($^{123}$I) channels are substantially fed from their higher charge isobars via $\beta^+$ decay and electron capture. The contribution of $\beta^+$ decay and electron capture has been calculated using the prescription of Cavinato $et$ $al.$\cite{cavinato1995study} and the independent cross-sections of these residues have been compared with PACE4 and EMPIRE calculations, which fairly reproduce the independent cross-sections of evaporation-residue within the experimental uncertainties. Interestingly, it has been observed that the $\alpha$-emitting channels, contrary to established findings in reactions involving $\alpha$ cluster projectiles (e.g., $^{12}$C, $^{16}$O, etc.) at the studied energy range, display negligible or no contribution of incomplete fusion (ICF) in $^{12}$C+$^{118}$Sn system. The absence of ICF has been verified through a complementary experiment in which the forward recoil ranges of $^{126}$Ba(4n) and $^{125}$Xe($\alpha$n) channels have been measured. Present measurements reveal anomalous suppression of ICF in the $^{12}$C+$^{118}$Sn system, providing new constraints on entrance-channel mass-asymmetry systematics absent in prior data. A comparison of ICF fraction as a function of entrance--channel mass--asymmetry for reactions involving $^{12}$C projectile with nearby targets indicates dissimilar behavior of the $^{118}$Sn target in $\approx$ 65-85 MeV energy range. 
\end{abstract}

\maketitle
\section{\label{sec:level1} Introduction} 
In heavy-ion (HI) induced reactions, the formation of the compound nucleus (CN) through complete fusion (CF) and its decay via light nuclear particles and/or characteristic $\gamma$-rays \cite{weisskopf1937statistics} generally dominates at near and above barrier energies, i.e., 3-7 MeV/nucleon \cite{trautmann1984dynamics,corradi2005excitation,parker1984complete}. In the case of CF, the entire projectile fuses with the target nucleus, involving all nucleonic degrees of freedom of interacting partners, forming a completely fused composite system (CS$^{*}$) \cite{lefort1976complete}. At this intermediate stage, pre-equilibrium emission might also occur with enhanced production cross-sections, particularly for lower $n$ channels \cite{sharma2015systematic,Manoj2024}. As the CS$^{*}$ attains thermal equilibrium, the excited CN may undergo fission, provided the favorable conditions for fission are met \cite{bohr1939mechanism,sierk1986macroscopic} or populate several evaporation residues (ERs) via emission of light nuclear particles and/or characteristic $\gamma$-rays \cite{weisskopf1937statistics}. 
Further, at these energies, the incomplete fusion (ICF) in which only a fraction of the projectile fuses with the target nucleus is found to compete with CF for higher $\ell$--values, i.e., $\ell>\ell_{crit}$, imparted into the system due to non-zero impact parameters in peripheral interactions \cite{singh2009role}. 

In the case of ICF, the incoming projectile is assumed to break up into $\alpha$-cluster(s) as it approaches the nuclear force field of the target nucleus \cite{Udagawa1980}. For example, $^{12}$C may breakup into $^{4}$He/$\alpha$ and $^{8}$Be/2$\alpha$ or $\alpha$+$\alpha$+$\alpha$ configurations. One of the fragments may get fused with the target nucleus, and the remnant is emitted in the forward cone with almost projectile velocity. As a result of fractional fusion of projectile with target nucleus; $(i)$ a reduced compound nucleus (RCN$^{*}$) is formed with less mass and charge as compared to the CF, $(ii)$ the recoil velocity of the reaction residues is expected to be less because of the fractional linear momentum transferred from projectile to target nucleus, and $(iii)$ the projectile-like fragments show maxima at forward angles \cite{nasirov2023new,barker1980direct,wilczynski1980incomplete, britt1961alpha,morgenstern1984influence,wilczynski1980incomplete,wu1980projectile,inamura1985gamma,singh2009probing,jha2020incomplete,amanuel2011investigation,tali2024comprehension}. The CN formed through CF and/or ICF (CN$^{*}$/RCN$^{*}$) leads to different ERs via emission of $x$n, p$x$n, $\alpha$$x$n/$\alpha$p$x$n, etc., channels. In some cases, it has been observed that the coupling of different population and decay routes leads to the formation of the same residues with enhanced production cross-sections. Such enhancements in the cross-sections have been attributed to pre-equilibrium emission, ICF, precursor decay, and other mechanisms, based on the observed characteristics of the reaction exit channels. Among these, precursor contribution arises from distinct decay modes, in which the evaporation residue decays into a daughter nucleus via emission of $\beta ^ +$/$\beta ^ -$, or electron capture (EC) \cite{cavinato1995study}. The $\beta$ decay \cite{coryell1953beta,rubio2023beta} proceeds via $(i)$ ${\beta}^-$ decays in neutron-rich nuclei and $(ii)$ $\beta^+$ decay in neutron-deficient nuclei with condition Q$_{re}$ $\geq$ 2m$_{e}c^2$ for $\beta^+$ decay; otherwise EC dominates via capture of an inner-shell electron. Hence, correcting the experimentally measured cross-sections to account for precursor decay contributions is essential to isolate the contribution of the primary reaction products, thereby ensuring a more precise determination of the independent production cross-sections of ERs. Therefore, evaluating the individual contribution of respective channels from the cumulative ones is crucial for better insights into the de-excitation patterns of CN formed in CF and/or ICF reactions \cite{cavinato1995study}. 

\begin{figure}
\includegraphics[scale=2.2]{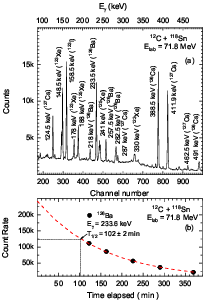}
\caption{\label{fig:spectraanddecaycurve}(a) A portion of $\gamma$ spectra of $^{12}$C+$^{118}$Sn system measured at E$_{lab}$ = 71.8 MeV. $\gamma$-lines associated with different ERs are marked. (b) The decay curve of $^{126}$Ba nuclei obtained by following 233.6 keV $\gamma$-line, indicating half-life of $\approx$102 min.}
\end{figure}

While prior studies involving $^{12}$C projectile with lighter mass targets, e.g., $^{93}$Nb \cite{agarwal2022role}, and $^{103}$Rh \cite{bindu2923}, reveal $>$10-20 \% ICF fraction at 5-7 MeV/nucleon, data for neutron-rich medium-mass targets, such as $^{118}$Sn, are scarce. A comprehensive understanding of nuclear reactions, including direct, compound nucleus, pre-equilibrium, and ICF processes, is hindered by limited experimental data in the medium-mass region. This measurement addresses a critical gap by measuring independent production cross-sections of neutron-deficient ERs (i.e., Cs, Xe, and I) produced in $^{12}$C+$^{118}$Sn system, enabling precise tests of precursor contributions in $\beta^+$/EC chains. In this work, the channel-by-channel excitation functions (EFs) of ERs have been measured in $^{118}$Sn($^{12}$C,\textit{x})$^{130}$Ba reaction at energies E$_{\textrm{lab}}$ $\approx$ 65-85 MeV to offer an extension of experiments performed with $^{12}$C (a 3-$\alpha$ configuration projectile) with $^{89}$Y \cite{chauhan2019measurement}, $^{93}$Nb \cite{agarwal2022role}, $^{103}$Rh \cite{bindu2923}, $^{115}$In \cite{mukherjee2006incomplete} and $^{128}$Te \cite{DiGregorio} targets. In these cases, ICF has been observed as a competing mode of reaction with CF and pre-compound emission at E$_{\textrm{lab}}$ $\approx$ 3-7 MeV/A. This paper is organized as follows: the experimental procedure to measure the absolute production cross-sections of different ERs is detailed in Sec. II. An attempt has been made to evaluate the $\beta ^ +$/EC contribution in the production of different ERs, and a unique case for $\lambda_p<<\lambda_d$ is also discussed. The analysis of EFs within the statistical model code PACE4 is detailed in Sec. III. Remarks on the unexpected absence of ICF in $^{12}$C+$^{118}$Sn reaction, and a complementary experiment to measure forward recoil ranges as a proof of linear momentum transferred from projectile to target nucleus are presented in sections IV and V of this article. Finally, Section VI discusses the summary and conclusions of this work.

\begin{table}
\caption{\label{tab:list of identified residues}List of identified evaporation residues produced in the $^{12}$C+$^{118}$Sn system, along with their spectroscopic properties \cite{nndc} is provided.}
\begin{ruledtabular}
\begin{tabular}{llll}
E$_{\gamma}$(keV) & I$_{\gamma}$($\%$) &  Residue & t$_{1/2}$  \\ \hline 
 
 233.6 & 19.67 $\pm$ 1.85 &  &   \\
 241.0 & 6.00 $\pm$ 0.68 &  $^{126}$Ba(4n) & 100 min\\
257.6 & 7.63 $\pm$ 0.72 &   &  \\ \hline 

124.7 & 11.38 $\pm$ 0.22 &  &  \\
412.0 & 62.90 $\pm$ 1.00 & $^{127}$Cs(p2n) & 375 min \\
462.3 & 5.08 $\pm$ 0.10 &  &   \\ \hline 

388.7 & 40.40 $\pm$ 2.00 &  &  \\
491.3 & 5.01 $\pm$ 0.44 & $^{126}$Cs(p3n) & 1.64 min  \\
925.2 & 4.52 $\pm$ 0.24 &  &   \\ \hline 

112.0   & 8.62 $\pm$ 0.83 &  &   \\
412.0   & 5.27 $\pm$ 0.51 & $^{125}$Cs(p4n) & 46.7 min \\
526.0   & 24.55$ \pm$ 2.38 &  &   \\ \hline 

188.4 & 53.80\footnotemark[1]  &  &   \\
243.4 & 29.97 $\pm$ 0.59 & $^{125}$Xe($\alpha n$) & 1014 min \\
453.8 & 4.67 $\pm$ 0.10 &  &   \\ \hline 

148.9 & 49.10 $\pm$ 0.60 &  &   \\
178.1 & 14.98 $\pm$ 0.76 & $^{123}$Xe($\alpha 3n$) & 123 min  \\ 
330.2 & 8.59 $\pm$ 0.50 &  &  \\ \hline 

148.6 & 2.62 $\pm$ 0.10 &  &   \\
350.1 & 7.80 $\pm$ 0.17 & $^{122}$Xe($\alpha 4n$) & 1206 min  \\
416.6 & 1.87 $\pm$ 0.04 &  &   \\ \hline 

159.0 & 83.60 $\pm$ 0.19 &  &   \\
440.0 & 0.39 $\pm$ 0.01 & $^{123}$I($\alpha p2n$) & 793 min   \\
529.0 & 1.27 $\pm$ 0.10 &  &   \\ \hline 

602.7 & 62.90 $\pm$ 0.71 &  &   \\
722.8 & 10.36 $\pm$ 0.12& $^{124}$I($\alpha pn$) & 4.176 days  \\
1691.0 & 11.15 $\pm$ 0.17 &  &   \\

\end{tabular}
\end{ruledtabular}
\footnotetext[1][error not mentioned in the reference]{}
\end{table}

\section{Experimental details}
The experiment was conducted using the 15-UD Pelletron Accelerator facilities at the Inter-University Accelerator Center (IUAC), New Delhi. Self-supporting Tin ($^{118}$Sn) targets (abundance = 24.22\%, enrichment = 99.96 $\%$) of thickness $\approx$ 0.27-0.34 mg/cm$^{2}$ fabricated using ultra-high vacuum deposition technique were used. Details of target fabrication are given elsewhere \cite{sood2020self}. Tin foils of dimensions 1.2 × 1.2 cm$^{2}$ were mounted onto aluminum target holders with a 1.0 cm diameter concentric hole and stacked with Al-catchers of sufficient thickness, $\approx$ 1.2-3 mg/cm$^{2}$, downstream of the targets to trap the evaporation residues recoiling out of the target foil. Two stacks of target-catcher foil assemblies were irradiated with $^{12}$C$^{+6}$ beam for 6.5-7.5 hrs, depending upon the half-life of interest, in the General Purpose Scattering Chamber (GPSC). The beam energy at the half thickness of different target foils is estimated to be 83.8, 78.6, 71.8, 70.8, and 64.8 MeV, achieved in two irradiations. The beam energy loss in targets and catcher foils is calculated using the SRIM code, based on range-energy correlation \cite{ziegler2010srim}. The beam current was continuously monitored and maintained at $\approx$ 1.5pnA during irradiation. The beam flux was calculated using the total charge collected in the Faraday cup installed at the beam dump.

\begin{figure}[h]
\includegraphics[scale=0.4]{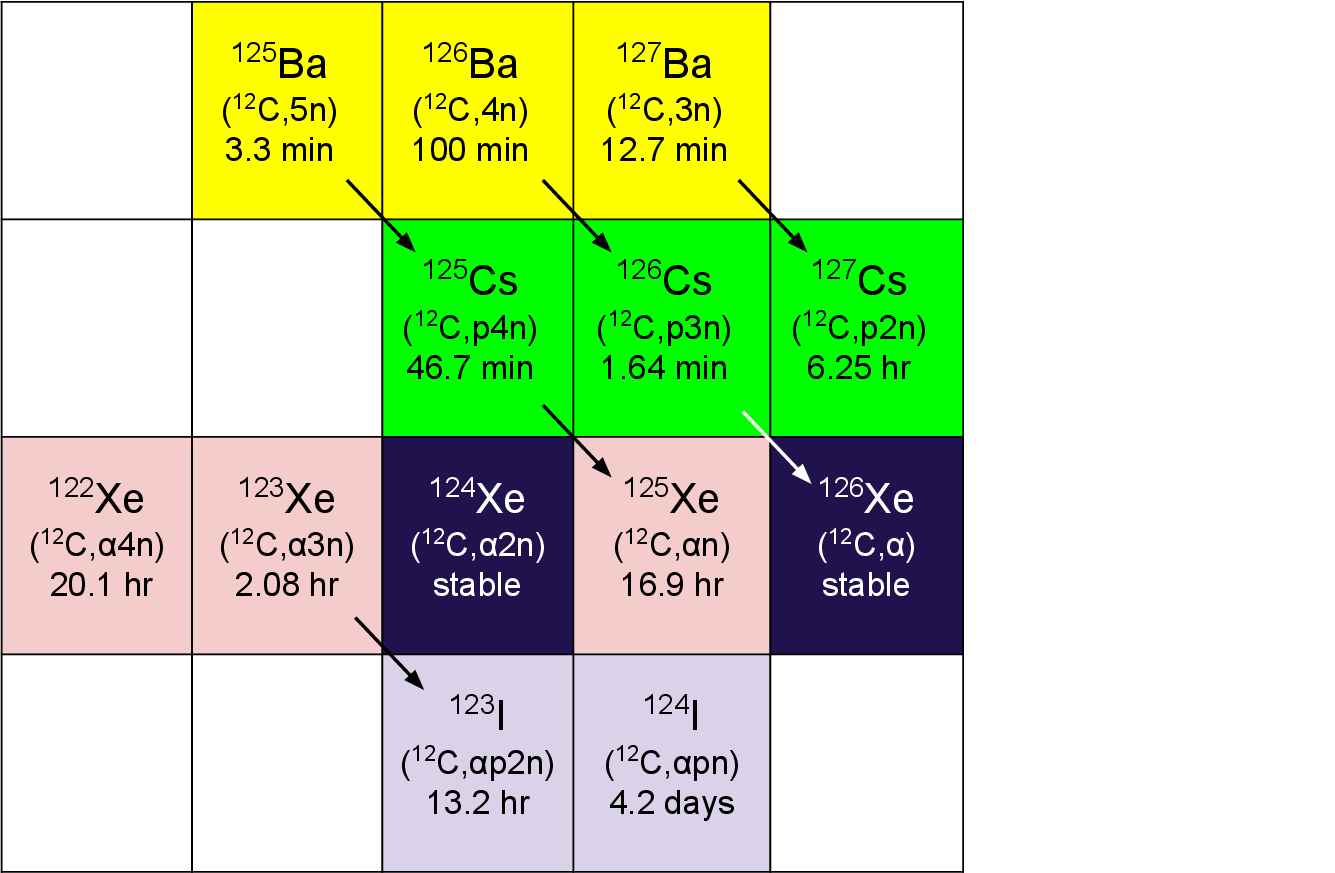}
\caption{\label{fig:decayresiduespicture}The decay chain of different evaporation residues is presented along with their half-lives \cite{nndc}. Nuclei that decay via  ${\beta}^{+}$ and/or electron capture (EC) are represented in different colors and are self-explanatory.}
\end{figure}

After irradiation, the target-catcher foil assemblies were removed from GPSC using an in-vacuum transfer facility for offline counting. The radioactivity produced during irradiations was counted offline using two pre-calibrated high-resolution n-type HPGe clover detectors with 470 cc active volumes, coupled to a CAMAC-based data-acquisition system. The resolution of the individual clover crystals and in the add-back mode was estimated to be $\approx$ 2-2.3 for 1332 keV $\gamma$-ray energy of $^{60}$Co source. The clover detectors may be operated in two distinct modes, i.e., $(i)$ by direct detection efficiency (G$ _{\epsilon D}$) estimated as the sum of four individual crystal efficiencies, and $(ii)$ by coincidence detection efficiency (G$ _{\epsilon C}$) also termed as add-back photo-peak efficiency. In the present work, the clover detectors were deployed in total detection mode, defined as the sum of the direct and coincidence detection efficiencies, to achieve the best performance of the composite detector \cite{duchene1999clover}. The detectors were calibrated using standard multi $\gamma$ sources, e.g., $^{60}$Co, $^{133}$Ba and $^{152}$Eu, of known strengths. 

\begin{figure}
\includegraphics[scale=1.3]{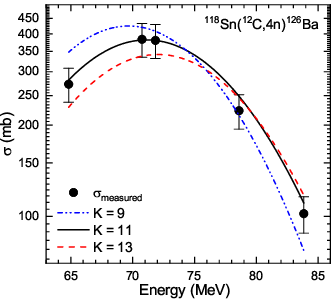}
\caption{\label{fig:4n channel}Experimentally measured excitation function of $^{126}$Ba compared with PACE4 calculations for different values of free parameter, i.e., K = 9,11,13. Lines and symbols are self-explanatory.}
\end{figure}

A representative $\gamma$ spectrum of $^{12}$C+$^{118}$Sn system obtained at E$_{\textrm{lab}}$ = 71.8 MeV is shown in Fig.\ref{fig:spectraanddecaycurve}(a). The characteristic decay $\gamma$-lines of $^{126}$Ba, $^{127}$Cs, $^{126}$Cs, $^{125}$Xe, $^{123}$Xe and $^{123}$I evaporation residues are marked in the spectrum. All evaporation residues were identified by following characteristic decay $\gamma$-lines and their half-lives obtained from the decay-curve analysis. Decay curve of $^{126}$Ba obtained by following 233.6 keV $\gamma$-line is presented in Fig.\ref{fig:spectraanddecaycurve}(b). As shown in the figure, the activity decreased by half over $\approx$102 $\pm$ 2 min, indicating the characteristic half-life of $^{126}$Ba nuclei. A similar procedure has been followed to analyze other reaction residues formed in this system. A list of identified evaporation residues is given in Table \ref{tab:list of identified residues} along with their spectroscopic properties.

\begin{table*}
\caption{\label{tab:cross-section cesium}Experimentally measured production cross-sections (in mb) of $^{126}$Ba and Cesium isotopes populated via CF reaction in $^{12}$C+$^{118}$Sn system.}
\begin{ruledtabular}
\begin{tabular}{llllllll}
  E$_{lab}$  &  $\sigma$ &  $\sigma_{cum}$  & $\sigma_{ind}$  &  $\sigma_{eff}$  & $\sigma_{ind}$  & $\sigma_{cum}$  & $\sigma_{ind}$  \\ 
 (MeV) &  ($^{126}$Ba) & ($^{127}$Cs) & $(^{127}$Cs) & $(^{126}$Cs) & $(^{126}$Cs) & $(^{125}$Cs) & $(^{125}$Cs) \\ \hline
 64.8 &   273$\pm$35   &  579$\pm$48  &  117$\pm$12  & 203$\pm$24  & 208$\pm$36   &  0.18$\pm$0.05  & 0.17$\pm$0.05  \\
 
 70.8 &   384$\pm$49   &  211$\pm$22  &   34$\pm$4   & 393$\pm$38  & 399$\pm$64   &  29$\pm$4       & 19$\pm$2.9         \\
 
 71.8 &   381$\pm$49   &  140$\pm$22  &   31$\pm$5   & 388$\pm$52  & 394$\pm$68   &  49$\pm$6       & 30$\pm$3.6       \\
 
 78.6 &   223$\pm$29   &   31$\pm$5   &   6$\pm$1  & 298$\pm$42  & 302$\pm$58   &  358$\pm$42     & 219$\pm$26        \\
 
 83.8 &   102$\pm$14   &   24$\pm$3   &   2$\pm$0.35 & 118$\pm$20    & 120$\pm$26   &  628$\pm$76     & 393$\pm$48  \\
\end{tabular}
\end{ruledtabular}
\end{table*}

Offline counting was conducted for 11 days to ensure that the long-lived evaporation residues were measured. The induced radioactivities have been used to estimate the production cross-sections of evaporation residues ($\sigma_{ER}$) at different energies using a standard formulation given elsewhere \cite{singh2008influence}. The cross-sections of $^{126}$Ba, $^{127, 126,125}$Cs, $^{125,123,122}$Xe and $^{124,123}$I residues are given in Table \ref{tab:cross-section cesium}, \ref{tab:cross-section xenon}, and \ref{tab:cross-section iodine} along with the experimental uncertainties. 

The errors in the experimental cross-sections of ERs may arise mainly because of: $(i)$ the non-uniform thickness of the samples, and an inaccurate estimate of foil thickness may lead to uncertainty in the determination of the number of target nuclei in the sample. To assess sample uniformity, their thicknesses were measured at multiple positions using the $\alpha$-transmission method. It is estimated that the sample material's thickness error is less than 1$\%$. $(ii)$ Fluctuations in the beam current may result in the variation of incident flux; proper care has been taken to keep the beam current constant, and hence, the error in flux has been avoided. $(iii)$ uncertainty in the determination of the geometry-dependent efficiency of the spectrometer. The error in the efficiency determination due to the statistical fluctuations in counts is estimated to be less than 2$\%$. $(iv)$ The dead time of the spectrometer was kept $\le$ 10$\%$ by suitably adjusting sample-detector distance. The measured error includes uncertainties in nuclear data, such as branching ratios and decay constants. The overall errors from these factors, including statistical errors, are estimated to be $\leq$ 15 $\%$ \cite{Siddiqueerror}.

\section{Data Analysis}
In heavy-ion induced reactions, the formation and decay of the excited compound nucleus may be described by the statistical model code PACE4 \cite{gavron1980statistical,singh2008influence}. In this code, the level density parameter, a = A/K MeV$^{-1}$, is a crucial input parameter that can be adjusted to reproduce the production cross-sections of evaporation residues populated via complete fusion. Here, A is the mass number of the compound nucleus, and K is a free parameter that can be optimized for a given projectile-target combination and excitation energy range. The channel-by-channel analysis of nine evaporation residues, i.e., $^{126}$Ba, $^{127, 126,125}$Cs, $^{125,123,122}$Xe, and $^{124,123}$I, identified in this work, has been performed in the framework of PACE4 and is presented in the following sub-sections. For ready reference, the decay chains of these residues are shown in Fig.\ref{fig:decayresiduespicture}, which represent the potential feeding patterns of different residues.

\subsection{$x$n channel: $^{126}$Ba residue}
In Fig.\ref{fig:4n channel}, experimentally measured excitation function of $^{126}$Ba residue populated via emission of four neutrons (4$n$) from the excited CN (i.e.,$^{130}$Ba) is compared with PACE4 calculations for K = 9,11,13. Lines through the data points represent the best fit with minimum residual error. It is evident from this figure that, for \enquote*{K = 11}, the PACE4 reproduces cross-sections of $^{126}$Ba residue reasonably well at the entire measured energy range, i.e., E$_{\textrm{lab}}$ $\approx$ 65-85 MeV, within the experimental uncertainties. The population of this residue may be attributed to the CF of $^{12}$C projectile with $^{118}$Sn target nuclei. Based on the analysis presented here, it may be inferred that the value of K = 11 is deemed optimized for the energy range E$_{\textrm{lab}}$ $\approx$ 65-85 MeV for $^{12}$C+$^{118}$Sn system. Therefore, the level density parameter a = A/11 MeV$^{-1}$ can be used to interpret other residues populated in this system.

\begin{figure*}
\includegraphics[scale=2.1]{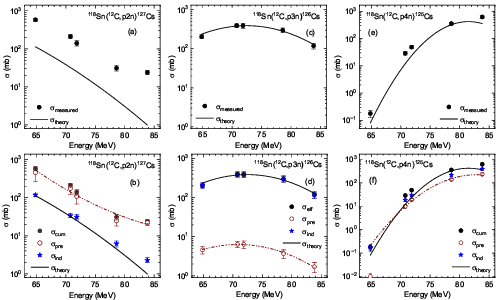}
\caption{\label{fig:matrixgraphpxn} Experimentally measured and theoretically calculated EFs of (a) $^{127}$Cs (p2n), (c) $^{126}$Cs (p3n), and (e) $^{125}$Cs (p4n) residues. In Figs. (b), (d), and (f), the values of cumulative, precursor, and independent cross-sections are compared with PACE4 calculations using a free parameter K = 11. Lines and symbols are self-explanatory.}
\end{figure*}

\subsection{p$x$n channels: $^{125,126,127}$Cs residues}
The EFs of three Cesium isotopes, i.e., $^{127}$Cs, $^{126}$Cs and $^{125}$Cs, populated via p$x$n channels (where $x$ = 2,3,4) have been measured in the present work and analyzed using theoretical model code PACE4 for optimized level density parameter, i.e., a = A/11 MeV$^{-1}$. Experimentally measured cumulative cross-sections for all p$x$n channels, along with the independent cross-sections of residues, are given in Table \ref{tab:cross-section cesium} and are plotted in Fig.\ref{fig:matrixgraphpxn}. As can be seen from Fig.\ref{fig:matrixgraphpxn}(a), (c), and (e), PACE4 substantially under/over-predicts the experimental EFs of $^{127}$Cs, $^{126}$Cs, and $^{125}$Cs residues, indicating the contribution from some physical effect which is not included in PACE4. In the decay chain of these residues, as given in Fig.\ref{fig:decayresiduespicture}, it has been noticed that the $^{127}$Cs, $^{126}$Cs, and $^{125}$Cs residues might have a contribution from their higher charge isobars following the $\beta^+$ decay and/or electron capture (EC). For example, $^{127}$Cs may be populated via emission of one proton and two neutrons, p$2$n channel, from the excited CN $^{130}$Ba$^*$ as,

\begin{eqnarray*}
{^{12}C + ^{118}Sn}  \rightarrow  {\left[^{130}Ba\right]}^* \rightarrow  {^{127}Cs + p2n};\\
{Q \approx -15.9 \, \textrm{MeV}} 
\end{eqnarray*}
and/or via EC/$\beta^+ (100\%)$ decay of $^{127}$Ba as,
\begin{eqnarray*}
{^{12}C + ^{118}Sn }   \rightarrow  {\left[^{130}Ba\right]}^*  \rightarrow  {^{127}Ba + 3n};
\end{eqnarray*}
\begin{eqnarray*}
{^{127}Ba(\epsilon + \beta^+})  \rightarrow  {^{127}Cs}.
\end{eqnarray*}

In such cases, if the half-life of the parent nuclei is smaller than the daughter nuclei, i.e., t$_{1/2}^{p}$ $\ll$ t$_{1/2}^{d}$, ($\lambda_d<<\lambda_p$), the independent cross-section of the daughter nuclei may be calculated by accounting for the precursor decay contribution as \cite{cavinato1995study},
\begin{eqnarray*}
{\sigma_{cum}^{d}}  =  {\sigma_{ind}^{d} + F^{p}\sigma_{cum}^{p}},
\end{eqnarray*}

where $\sigma_{cum}^{d}$ and $\sigma_{ind}^{d}$ are the cumulative and independent cross-sections of the daughter nuclei, and $\sigma_{cum}^{p}$ represents the cross-section of the precursor. F$^{p}$ is the fraction of precursor decay that depends upon the branching ratio (P$^{p}$) of the precursor to daughter nuclei and is defined as,
\begin{eqnarray*}
{F^{p}} =  P^{p}\frac{\lambda_p}{\lambda_p - \lambda_d} = P^{p}\frac{t_{1/2}^{d}}{t_{1/2}^{d} - t_{1/2}^{p}},
\end{eqnarray*}

where, t$_{1/2}^{d}$ and t$_{1/2}^{p}$ are the half-lives of daughter and precursor nuclei. Following the prescription given in ref. \cite{cavinato1995study}, the independent cross-sections of the daughter nuclei have been calculated from the experimentally measured cumulative cross-section as,
\begin{eqnarray}
\label{precursorcontribution}
{\sigma_{ind}^{d} = {\sigma_{cum}^{d}} - {P^{p}\frac{t_{1/2}^{d}}{t_{1/2}^{d} - t_{1/2}^{p}}}\sigma_{cum}^{p}}.
\end{eqnarray}

The values of branching ratios and the half-lives required for obtaining the precursor decay coefficient, F$^{p}$ in the case of $^{127}$Cs, have been taken from Refs.\cite{AHashizume}. The value of $\sigma_{ind}$ for the formation of $^{127}$Cs residue is evaluated as,
\begin{eqnarray*}
{\sigma_{ind}(^{127}Cs)}  =  {\sigma_{cum}(^{127}Cs) - 1.04*\sigma_{cum}(^{127}Ba)},
\end{eqnarray*}
where $\sigma_{cum}(^{127}$Ba) is the yield of precursor.

In Fig.\ref{fig:matrixgraphpxn}(b), the precursor contribution ($\sigma_{pre}$) along with $\sigma_{cum}$ and $\sigma_{ind}$ of $^{127}$Cs residue is compared with PACE4 calculations for $^{127}$Cs residue. As can be seen in this figure, the value of ${\sigma_{ind}}$ ($^{127}$Cs) at all measured energies is found to be in good agreement with PACE4 calculations, indicating the population of $^{127}$Cs via the p$2n$ channel in a CF reaction. A similar analysis has been conducted for $^{125}$Cs, $^{125}$Xe and $^{123}$I residues, given that t$_{1/2}^{p}$ $\leq$ t$_{1/2}^{d}$, respectively.

Further, as indicated in Fig.\ref{fig:decayresiduespicture}, $^{126}$Cs is expected to be populated via $(i)$ emission of one proton and three neutrons from $^{130}$Ba$^*$, and $(ii)$ electron capture (98.3$\%)$ / $\beta^+ (1.7\%)$ decay \cite{reus1983atomic, Iimura}. In this case, the half-life of the higher-charge precursor is 100 min, which is much longer than that of the daughter nucleus, $^{126}$Cs (1.64 min). Therefore, the precursor contribution in this case can be deduced as \cite{cavinato1995study},

\begin{eqnarray*}
{N_d (t)}  =  {C_{t=0}e^{-\lambda_{d}t^{'}}} + {P^{p}\frac{\lambda_{p}}{\lambda_d - \lambda_p}N_p(T) e^{-\lambda_p t^{'}}},
\end{eqnarray*}
where N$_d (t)$ and N$_p (T)$ are the number of daughter nuclei produced at time `t$^{'}$= t-T' and the number of precursor nuclei produced at time irradiation stopped, C$_{t=0}$ is the cumulative (precursor + daughter) number of nuclei at the end of irradiation. 

Now as $\lambda_p << \lambda_d$, above equation becomes,
\begin{eqnarray*}
{N_d (t)}  =  {C_{t=0}e^{-\lambda_{d}t^{'}}} + {P^{p}\frac{\lambda_{p}}{\lambda_d}N_p(T) e^{-\lambda_p t^{'}}},
\end{eqnarray*}
As, $\lambda_p << \lambda_d$ and t $\rightarrow{\infty}$ , t$^{'}$  = t - T $\rightarrow{\infty}$, first term becomes negligible as compared to second term and 

\begin{eqnarray*}
{N_d (t)}  \rightarrow{{P^{p}\frac{\lambda_p}{\lambda_d}N_p(T) e^{-\lambda_p t}}  },
\end{eqnarray*}

Similarly, 

\begin{eqnarray*}
{\sigma_d }  =  {\sigma_{eff} + {P^p\frac{\lambda_p}{\lambda_d}{\sigma_p}}},
\end{eqnarray*}
 
where $\sigma_{eff}$ is the effective cross-section that we got from the decay curve of $^{126}$Cs residue, giving the half-life of precursor nuclei according to $N_d(t)$.

Further, $^{125}$Cs is expected to be populated via p4n channel and/or via decay of its higher charge isobar in a reaction type, $^{12}$C($^{118}$Sn,5n)$^{125}$Ba, where $^{125}$Ba decays into $^{125}$Cs via $\beta^+$/EC (100$\%)$ \cite{Katakura} emission as indicated in Fig.\ref{fig:decayresiduespicture}. In this case, the half-life of $^{125}$Cs (46.7 min) is longer than that of its precursor $^{125}$Ba (3.3 min). The independent production cross-section of $^{125}$Cs residue may be calculated as,

\begin{eqnarray*}
{\sigma_{ind}(^{125}Cs)} = {\sigma_{cum}(^{125}Cs) - 1.04*\sigma_{cum}(^{125}Ba)}.
\end{eqnarray*}

As the cross-section of $^{125}$Ba has not been measured in this work, the cross-section of $^{125}$Ba is taken from PACE4 (using K=11) to reduce precursor contribution from the cumulative yield of $^{125}$Cs. The experimentally measured and systematically deduced independent cross-sections of $^{126,125}$Cs residues at different energies are compared with PACE4 calculations in Fig.\ref{fig:matrixgraphpxn}(d) and (f). As can be seen from these figures, the values of independent cross-sections of $^{126,125}$Cs residues at all energies are in good agreement with PACE4 calculations within the experimental uncertainties, indicating the population of these residues via CF of $^{12}$C with $^{118}$Sn. 

\begin{figure}
\includegraphics[scale=2.6]{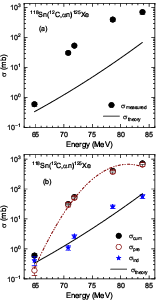}
\caption{\label{fig:alpha-n channel}$(a)$Experimentally measured cross-sections of $^{125}$Xe($\alpha$n), and $(b)$ the independent cross-sections of $^{125}$Xe residue compared with PACE4 calculations corresponding to free parameter K = 11.}
\end{figure}

\subsection{\label{sec:level3}$\alpha x$n channels: $^{125,124,123}$Xe residues}
In this work, three isotopes of xenon, i.e., $^{125}$Xe, $^{123}$Xe, and $^{122}$Xe, have been identified. The excitation function of $^{125}$Xe ($\alpha$n) is compared with PACE4 as shown in Fig.\ref{fig:alpha-n channel}(a). As shown in this figure, the PACE4 underestimates the experimental cross-sections at all energies, indicating a substantial contribution from processes other than CF. As per the existing understanding of low energy nuclear reactions involving $\alpha$-cluster structure projectiles, e.g., $^{12}$C, $^{16}$O, $^{20}$Ne, etc., $^{125}$Xe, $^{123}$Xe and $^{122}$Xe residues may be populated via three modes, $(i)$ complete fusion, $(ii)$ incomplete fusion \cite{bindu2923,mukherjee2006incomplete,kamalkumar}, and/or $(iii)$ precursor decay/EC. The complete fusion of $^{12}$C with $^{118}$Sn may lead to the formation of excited $^{130}$Ba$^*$, which may eventually decay via light nuclear particles. For example, $^{125}$Xe may be formed via the emission of an $\alpha$ particle and a neutron and/or by emitting two protons and three neutrons from excited $^{130}$Ba$^*$, as,

\begin{eqnarray*}
{^{12}C + ^{118}Sn }   \rightarrow  {\left[^{130}Ba\right]}^*  \rightarrow  {^{125}Xe + \alpha n};\\
{Q \approx -10.6 \, \textrm{MeV}}, and/or
\end{eqnarray*}
\begin{eqnarray*}
{^{12}C + ^{118}Sn }   \rightarrow  {\left[^{130}Ba\right]}^*  \rightarrow  {^{125}Xe + 2p3n}.
\end{eqnarray*}

\begin{figure}
\includegraphics[scale=1.25]{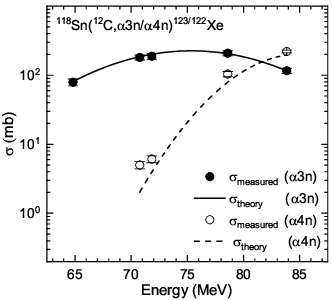}
\caption{\label{fig:apha3nalpha4n}Experimentally measured EFs of $^{123}$Xe($\alpha$3n) and $^{124}$Xe($\alpha$4n) residues compared with PACE4 predictions for free parameter, K = 11.}
\end{figure}

Since $^{12}$C is a 3-$\alpha$ cluster projectile, it can break into $^{8}$Be and $^{4}$He which may give rise to $^{125}$Xe via fusion of $^{8}$Be during the reaction. For example, $^{8}$Be may fuse with $^{118}$Sn to form an RCN (i.e., $^{126}$Xe$^*$), and the remnant $\alpha$ is released with beam velocity in the forward cone as a spectator. The $^{126}$Xe$^*$ may further emit a neutron to form $^{125}$Xe as,
\begin{eqnarray*}
{^{12}C [^{8}Be + ^{4}He] }   \rightarrow  {\left[^{126}Xe\right]}^* + ^{4}He \rightarrow  {^{125}Xe + n};\\
{Q \approx -10 \, \textrm{MeV}}.
\end{eqnarray*}
Further, $^{125}$Xe may also have contribution from its higher charge isobar $^{125}$Cs via electron capture (60\%) or via $\beta^+ (40\%)$ decay \cite{reus1983atomic,Katakura} as,
\begin{eqnarray*}
{^{12}C + ^{118}Sn }   \rightarrow  {\left[^{130}Ba\right]}^*  \rightarrow  {^{125}Cs + p4n};
\end{eqnarray*}
\begin{eqnarray*}
{^{125}Cs(\epsilon + \beta^+})  \rightarrow  {^{125}Xe}.
\end{eqnarray*}
In this case, the half-life of $^{125}$Xe is 1014 min whereas the half-life of precursor $^{125}$Cs is 46.7 min (i.e., t$_{1/2}^{p}$ $\ll$ t$_{1/2}^{d}$). Therefore, the independent cross-section of $^{125}$Xe may be calculated as, 
\begin{eqnarray*}\label{125CsTO125Xe}
{\sigma_{ind}(^{125}Xe)}  =  {\sigma_{cum}(^{125}Xe) - 1.04*\sigma_{cum}(^{125}Cs)},
\end{eqnarray*}
The independent cross-section of $^{125}$Xe is compared with PACE4 calculations in Fig.\ref{fig:alpha-n channel}(b). As seen from this figure, the independent cross-section of $^{125}$Xe agrees reasonably well with the theoretical calculations, indicating the population of this residue solely via complete fusion despite being an $\alpha$-emitting channel.

\begin{table}
\caption{\label{tab:cross-section xenon}Experimental cross-sections of xenon isotopes populated via $\alpha$xn channels.}
\begin{ruledtabular}
\begin{tabular}{lllll}
E$_{lab}$  &  $\sigma_{cum}(^{125}$Xe)  & $\sigma_{ind}(^{125}$Xe)  &  $\sigma(^{123}$Xe) & $\sigma(^{122}$Xe)   \\ 
 (MeV) &(mb) & (mb) & (mb) & (mb) \\ \hline
 64.8 &   0.6$\pm$0.09 &  0.41$\pm$0.13   & 79$\pm$9     &    -   \\
 70.8 &   32$\pm$2   &   1.1$\pm$0.2  & 181$\pm$15   & 5$\pm$0.6  \\
 71.8 &   54$\pm$4    &   2.6$\pm$0.4  & 189$\pm$18   & 6.1$\pm$0.7   \\
 78.6 &   401$\pm$30  &    25.7$\pm$3.6  & 209$\pm$18   & 104$\pm$10   \\
 83.8 &   715$\pm$55 &    56.7$\pm$8.1  & 116$\pm$10   & 220$\pm$10   \\
\end{tabular}
\end{ruledtabular}
\end{table}

A similar analysis has been performed to calculate the independent cross-sections of $^{123}$Xe and $^{122}$Xe residues. In Fig.\ref{fig:apha3nalpha4n}, the experimentally measured independent cross-sections of $^{123}$Xe and $^{122}$Xe residues are compared with PACE4 calculations. The cross-sections of $^{123}$Xe and $^{122}$Xe agree with the PACE4 calculations using the free parameter K = 11. Therefore, it may be safely inferred that the $^{123}$Xe and $^{122}$Xe residues are populated via complete fusion through $\alpha 3n $ and $\alpha 4n$ channels.

\subsection{\label{sec:level4} $\alpha$p$x$n channels: $^{124,123}$I residues} 
\begin{figure}
\includegraphics[scale=1.25]{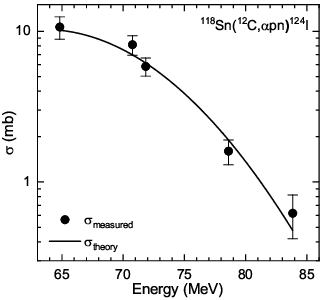}
\caption{\label{fig:alpha-pn channel}Experimentally measured EFs of $^{124}$I ($\alpha$pn) in comparison with PACE4 predictions corresponding to free parameter K = 11.}
\end{figure}

\begin{figure}
\includegraphics[scale=2.6]{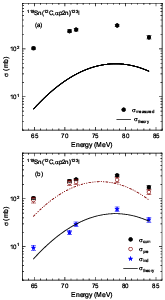}
\caption{\label{fig:alpha-p2n channel}Experimentally measured excitation function of (a) $^{123}$I ($\alpha$p2n) is compared with PACE4 calculations for free parameter K = 11. (b) compares the cumulative, precursor, and independent cross-sections with PACE4 calculations. Lines and symbols are self-explanatory.}
\end{figure}

The EFs of Iodine residues, i.e., $^{124}$I and $^{123}$I, are presented in Fig.\ref{fig:alpha-pn channel} and \ref{fig:alpha-p2n channel} along with PACE4 calculations. The residue $^{124}$I may be populated via $\alpha$pn channel (Q $\approx-$17.6 MeV) or emission of three protons and three neutrons from the excited CN$^*$ ($^{130}$Ba$^*$) in a CF reaction. This residue may also be populated via ICF of $^{8}$Be from $^{12}$C by emitting one proton and one neutron from the reduced compound nucleus $^{126}$Xe$^*$ (Q $\approx-$17.1 MeV), while an $\alpha$-particle behaves as a spectator. As seen in Fig. \ref{fig:alpha-pn channel}, the experimentally measured cross-sections of $^{124}$I agree reasonably well with PACE4 calculations, indicating that this residue is populated via CF only. 

Further, the experimental EF of $^{123}$I is plotted in Fig.\ref{fig:alpha-p2n channel}(a) with PACE4 calculations. As shown in this figure, the PACE4 consistently underestimates the experimental cross-sections for the $^{123}$I residue at all measured energies. This enhancement of experimental cross-sections over the PACE4 calculations may be due to the contribution of ICF and/or the contribution of precursor nuclei $^{123}$Xe through electron capture (77.5\%) or $\beta^+ (22.5\%)$ \cite{reus1983atomic,junchen}. In this case, the half-life of $^{123}$I is 783 min, which is significantly higher than the half-life of the higher charge precursor $^{123}$Xe (i.e., t$_{1/2}$ = 123 min). Therefore, the independent cross-section of $^{123}$I is estimated as,
\begin{eqnarray}\label{123Xeto123I}
{\sigma_{ind}(^{123}I)}  =  {\sigma_{cum}(^{123}I) - 1.19*\sigma_{cum}(^{123}Xe)}.
\end{eqnarray}

\begin{table}
\caption{\label{tab:cross-section iodine}Experimental cross-sections of Iodine residues populated via CF and/or ICF in $^{12}$C+$^{118}$Sn system.}
\begin{ruledtabular}
\begin{tabular}{llll}
E$_{lab}$  &  $\sigma(^{124}$I)  & $\sigma_{cum}(^{123}$I)  &  $\sigma_{ind}(^{123}$I)   \\ 
 (MeV) & (mb) & (mb) & (mb)  \\ \hline
 64.8 &   10.7$\pm$1.8  &  103$\pm$6     & 9.5$\pm$1.2   \\
 70.8 &   8.1$\pm$1.2   &   234$\pm$16   & 20$\pm$2.1    \\
 71.8 &   5.8$\pm$0.8   &   253$\pm$16     & 29$\pm$3.4   \\
 78.6 &   1.6$\pm$0.3 &    308$\pm$18      &  61$\pm$6.3   \\
 83.8 &   0.6$\pm$0.2   &    174$\pm$14    & 37$\pm$4.3    \\
\end{tabular}
\end{ruledtabular}
\end{table}

The resulting independent cross-section of $^{123}$I at different energies is plotted with PACE4 calculations in Fig. \ref{fig:alpha-p2n channel}(b). As can be seen from this figure, the independent cross-section of $^{123}$I residue agrees reasonably well with the theoretical calculations within the experimental uncertainties, indicating the population of $^{123}$I residue only via the CF through the emission of an $\alpha$, a proton, and two neutrons or via emission of three protons and four neutrons from the excited CN$^*$ $^{130}$Ba$^*$ (Q $\approx-$16.6 MeV). 

\begin{figure}
\includegraphics[scale=1.3]{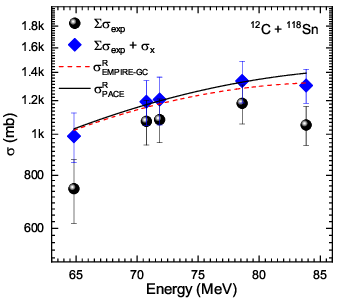}
\caption{\label{fig:PaceEmpireExperimentalDeduced}A comparison of experimental total x-sec with PACE4 and EMPIRE calculations for free parameter K = 11 in the case of PACE4 and LEVDEN = 2 for EMPIRE using Gilbert-Cameron level densities. Lines and symbols are self-explanatory.}
\end{figure}

\section{Insights from statistical model comparisons}

The experimentally measured cross-sections of ERs have been validated against PACE4 and EMPIRE predictions to confirm consistency with existing statistical evaporation models, thereby allowing the isolation of non-statistical effects, such as precursor decays and/or electron capture. The analysis shows that the experimentally measured EFs of $x$n channels agree reasonably well with the theoretical ones within the experimental uncertainty of 10-15\% ($\chi^2$$\approx$1.2), validating the methodology adopted in this work. However, the discrepancies in pxn/$\alpha$$x$n channels have been attributed to the feeding from $\beta^+$/EC (e.g., 50-70$\%$ for $^{125}$Cs from $^{125}$Ba residues) processes.

For better insights into how PACE4 calculations with optimized input parameter, K = 11, explain the sum of the independent cross-section of all ERs ($\sum \sigma _ {exp}$) measured in this work, the value of $\sum \sigma _ {exp}+\sigma _ {x}$, corrected with the missing cross-section ($\sigma _ {x}$ ) of channels which could not be measured (i.e., 3n, 5n, $\alpha$ and $\alpha$2n), is plotted in Fig.\ref{fig:PaceEmpireExperimentalDeduced}. In this figure, the predictions of EMPIRE3.2.3 (LEVDEN = 2) \cite{herman2007empire,capote2007empire} are also compared along with the PACE4 calculations. As shown in this figure, experimentally measured cross-sections of ERs, corrected for missing channels, agree reasonably well with PACE4 and EMPIRE-GC calculations within the experimental uncertainties of 1-7$\%$. Based on the excellent agreement between experimental data and theoretical calculations, it can be safely inferred that all the residues studied in this work are produced solely via CN formation and its decay in a CF reaction.

\section{Remarks on the unexpected absence of incomplete fusion}

In some of the recent investigations, contrary to the well-established onset of ICF at E/A $\geq$ 10 MeV \cite{arnell1983incomplete}, a significant fraction of ICF has been reported at energies as low as E/A $\leq$ 4-7 MeV and even at slightly above the Coulomb barrier  \cite{bindu2923,mukherjee2006incomplete,yadav2017systematic,sharma2014influence,kumar2018sensitivity,DiGregorio,kumar2003complete,agarwal2022role,yadav2023understanding,sagwal2023unfolding, kamalkumar, gollan2021one,chakrabarty2000complete}. Moreover, Jha et al. \cite{jha2020incomplete} presented a systematic study of low-energy ICF across a diverse spectrum of projectile-target combinations. It has been reported that the ascertained trends indicate a noticeable dependency of ICF on the specific characteristics of the projectile, notably on its Q-$\alpha$ value ($^{12}$C, Q-$\alpha$ $\approx$ -7.37 MeV). Bindu $et$ $al$.\cite{bindu2923}, observed significant ICF for $^{12}$C+$^{103}$Rh system, Mukherjee $et$ $al$.\cite{mukherjee2006incomplete}, for $^{12}$C+$^{115}$In system, and Digregorio $et$ $al$.\cite{DiGregorio}, for $^{12}$C+$^{128}$Te in the same mass region involving $^{12}$C projectile. Generally, with $\alpha$-cluster structure projectiles, such as $^{12}$C, $^{16}$O, $^{20}$Ne, etc., the ICF onsets at lower energies and starts competing with CF as the energy increases. However, there appears to be no or negligible contribution from the ICF process in $^{12}$C+$^{118}$Sn system at the measured energies, unlike the established findings in $^{12}$C-induced reactions, refer to Fig. \ref{fig:alpha-n channel}-\ref{fig:alpha-p2n channel}. 

\begin{figure}
\includegraphics[scale=1.3]{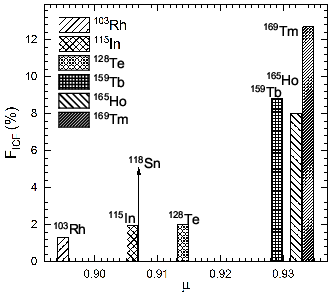}
\caption{\label{fig:FICF} A comparison of F$_{\textrm{ICF}}$ values derived at a fixed relative velocity ($v_{rel}$ = 0.053) as a function of entrance channel mass asymmetry ($\mu$) for various target nuclei, $^{103}$Rh \cite{bindu2923}, $^{115}$In  \cite{mukherjee2006incomplete}, $^{128}$Te \cite{DiGregorio}, $^{159}$Tb \cite{yadav2012large}, $^{165}$Ho \cite{gupta2000complete} and $^{169}$Tm \cite{chakrabarty2000complete} involving $^{12}$C as projectile.}
\end{figure}

To probe how strength of ICF fair for medium mass target nuclei, around $^{118}$Sn mass region, involving $^{12}$C projectile, the ICF fraction (F$_{\textrm{ICF}}$) has been analyzed for different projectile-target combinations at a constant relative velocity, i.e., v$_{\textrm{rel}}$ = 0.053. The value of F$_{\textrm{ICF}}$ is plotted as a function of entrance-channel mass-asymmetry ($\mu$) in Fig.\ref{fig:FICF}. As can be seen from the distribution of F$_{\textrm{ICF}}$, the contribution of ICF for the given targets increases from $\approx$8 \% to 12 \% with the entrance-channel mass-asymmetry. However, for the $^{118}$Sn target (studied in the present work), the contribution of ICF is zero or negligible. The current findings reinforce the target dependence on ICF processes. Unlike existing measurements involving $^{12}$C projectile where ICF contributes 10-18$\%$ \cite{nasirov2023new,barker1980direct,wilczynski1980incomplete, britt1961alpha,morgenstern1984influence,wilczynski1980incomplete,wu1980projectile,inamura1985gamma,singh2009probing,jha2020incomplete,amanuel2011investigation,tali2024comprehension}, no enhancement in $\alpha x$n channels suggests higher neutron excess in $^{118}$Sn (e.g., N/Z=1.41 vs. 1.35 for $^{103}$Rh) suppressing F$_{ICF}$ by $>$15$\%$. This work indicates a deviation from Morgenstern's mass-asymmetry systematics\cite{morgenstern1984influence}, hinting towards the neutron-skin thickness (t$_n$ $\approx$ 0.18 fm from PREX) effect as a key parameter.

\section{Forward-recoil ranges: Proof of fractional linear momentum transfer in $^{12}$C+$^{118}$Sn system}

As a complementary measurement, the presence of ICF, if any, can be investigated through the measurement of forward ranges of recoils (FRRs), which depend on the degree of linear momentum transferred ($\rho_{LMT}$) from projectile to target nuclei \cite{ShuaibRRD,MahatoRRD,AsnainRRD}. The $\rho_{LMT}$ is defined as ${P_{frac}}/{P_{proj}}$, where ${P}_{frac}$ is the linear momentum fraction of the fused projectile component, and ${P}_{proj}$ is the entire linear momentum of the projectile. For the maximum value of $\rho_{LMT}$, the entire projectile fuses with the target nucleus, the CN achieves the highest range in the stopping medium. However, in the case of ICF, a partial $\rho_{LMT}$ transfer leads to the formation of an incompletely fused composite system for which quantities such as mass, energy, and CN momentum will be less than those of CF. The ERs populated via a less LMT display a relatively smaller range in the stopping medium than that of the entire LMT population. Therefore, the experimentally measured FRRs can be used to investigate complete and/or incomplete fusion in a reaction.

\begin{figure}[h]
\includegraphics[scale=2.5]{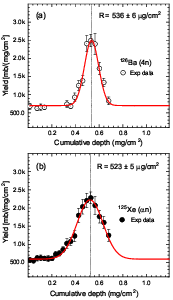}
\caption{\label{fig:4nalphanRRD} Experimentally measured FRRs of (a) $^{126}$Ba (4n) and (b) $^{125}$Xe ($\alpha$n) residues measured in $^{12}$C+$^{118}$Sn reaction at beam energy of 78.9 MeV.}
\end{figure}

To strengthen the findings of this work, the FRRs associated with $\alpha$ and non-$\alpha$ emitting channels in $^{12}$C+$^{118}$Sn system, expected to be populated via ICF and CF, respectively, have been measured using the same setup and method described elsewhere \cite{siddiqueRRD}. However, a brief account of unique experimental conditions is given here. A stack of a self-supporting Tin ($^{118}$Sn) target and several thin Al-catcher foils (thickness $\approx$ 0.03 - 0.04 mg/cm$^{2}$) - mounted downstream of the target, was irradiated with a $^{12}$C beam of 78.9 MeV for a duration of approximately 24 hours at a beam current $\approx$ 1.35 pnA. The Al catcher foils were mounted behind the target to trap the reaction residues recoiling from the target foil during irradiation. After irradiation, the activity induced in each Al-catcher foil was measured offline using two pre-calibrated high-resolution HPGe clover detectors with an active volume of 470 cc. The resolution of the detectors was estimated to be $\approx$ 2 keV for the 1.33 MeV $\gamma$-ray of $^{60}$Co before and after the counting. The residue identification and cross-section calculation procedure is similar to the EFs measurement.

Fig.\ref{fig:4nalphanRRD}(a)-(b) show FRRs of 4n and $\alpha$n emitting channels, respectively, where the normalized yield is plotted as a function of cumulative thickness of Al-catcher foils. As shown in these figures, the FRRs of both channels can be fit by a Gaussian distribution, indicating that a single linear-momentum-transfer component is responsible for forming these residues. The recoil range distribution of $^{126}$Ba -- populated via emission of 4 neutrons from the CN $^{130}$Ba -- is peaking at most probable range R = 536$\pm$6$\mu$g/cm$^{2}$, and the residue $^{125}$Xe -- populated via emission of an $\alpha$-particle and a neutron or 2p3n (2 protons and 3 neutrons) -- is showing a range of R = 525$\pm$5$\mu$g/cm$^{2}$. As both the residues show the same most probable range (i.e, R = 525-536$\mu$g/cm$^{2}$) within the experimental uncertainties, refer to Fig.\ref{fig:4nalphanRRD}(a)-(b), it can be inferred that both the residues are populated via CF dynamics attributed to the entire linear momentum transfer from projectile to target nucleus. It may be noted that the $^{125}$Xe residue is expected to be populated via ICF, based on numerous studies in the literature for similar projectile-target systems, with a substantially smaller range than CF residues. However, in this case, the FFR data suggest a single linear-momentum-transfer component peaking at approximately the same most probable range in the stopping medium. This confirms that both residues are predominantly produced via CF of $^{12}$C with $^{118}$Sn, with negligible or no contribution from ICF reactions. Further, the recoil-range distribution data reiterate the findings of this work, ruling out the presence of ICF, as only the entire LMT component associated with CF of the projectile has been observed. This highlights the target-specific reaction dynamics - first observed in $^{12}$C+$^{118}$Sn system at the studied energies.

\section{Summary and conclusions}
To summerise, channel-by-channel EFs of nine evaporation residues, i.e., $^{126}$Ba, $^{127, 126,125}$Cs, $^{125,123,122}$Xe and $^{124,123}$I, produced in $^{12}$C+$^{118}$Sn system have been measured at energies $\approx$ 64-84 MeV and analyzed within the framework of theoretical model codes PACE4 and EMPIRE3. The ERs studied in this work are neutron-deficient and hence predominantly decayed via ${\beta}^{+}$/EC. An effort has been made to estimate the independent production cross-sections of different ERs by analyzing the contributions of cumulative and precursor decays to their formation. Remarkably, a substantial contribution of ${\beta}^{+}$/EC has been observed in p$2$n(80-90$\%$), p$4$n(35-40$\%$), $\alpha$n($\approx$90$\%$) and $\alpha$p$2$n(75-95$\%$) channels. The precursor corrections reduce $\sigma_{ind}$\ by 40-60$\%$ in p$x$n channels, enabling accurate reaction rate estimation for astrophysical processes.

Analysis of experimental EFs of different ERs indicates the dominance of CF in the $^{12}$C+$^{118}$Sn system, whereas previous investigations involving $^{12}$C projectiles suggested a competing presence of ICF. This work reveals an irregular behavior in the $^{12}$C+$^{118}$Sn system compared with similar systems, which may be attributed to the target-nuclei structure. The existence of ICF in $^{12}$C-induced reactions is not universal across the studied energy range, underscoring the need for system-dependent investigations. Further, this work emphasizes the methodological necessity of extracting independent production cross-sections rather than relying on cumulative yields of ERs, as standard statistical model codes do not account for precursor feeding contributions within cumulative cross-sections. The current analysis addresses a precursor-decay scenario in which the parent nucleus has a significantly longer half-life than its daughter nuclei. The successful extraction of the independent cross-sections in this case validates the reliability and robustness of the adopted evaluation methodology for complex decay chains.

The contribution of ICF is found to be zero or negligible in $^{12}$C+$^{118}$Sn system, contrary to reactions involving the targets in the same mass region. To support the findings of the EFs measurement, forward recoil ranges (FRRs) of ERs have been measured in a complementary experiment. Analysis of FRRs of representative 4n and $\alpha$n emission channels reveals single Gaussian-shaped distributions for both channels, indicating the presence of a single linear momentum transfer component in the formation of $^{126}$Ba and $^{125}$Xe residues, confirming the absence of ICF in the $^{12}$C+$^{118}$Sn system at the measured energies. The unexpected absence of ICF in $\alpha$-emitting channels challenges existing projectile breakup models, possibly indicating the influence of neutron-skin thickness effects that suppress peripheral fusion by $\approx15-20\%$ compared to nearby systems. The anomalous absence ICF provides new evidence for neutron-rich target effects in low-energy nuclear reactions. These findings may help refine reaction models for producing exotic nuclei at facilities such as RIBF or FRIB. To better understand these aspects, more inclusive experiments may be performed using a broader range of Tin targets with $\alpha$-cluster projectiles at low incident energies.  

\begin{acknowledgments}
The authors thank the Inter-University Accelerator Center, New Delhi, for providing the experimental facilities for these measurements. One of the authors, Priyanka, acknowledges the Department of Science $\&$ Technology (DST), Govt. of India, for the doctoral fellowship in an Indo-Russian Project with Ref. No. DST/INT/RUS/RSF/P-23, and the National Mission on Interdisciplinary Cyber-Physical Systems (NM-ICPS) for financial support through the Technology Innovation Hub, iHub - AWaDH, established at IIT Ropar. 
\end{acknowledgments}

\nocite{*}
\bibliography{apssamp}

\end{document}